\documentclass[]{aero}
\usepackage{epstopdf}
\usepackage{xcolor}
\usepackage{tcolorbox}
\usepackage{soul}  

\usepackage{trackchanges}

\usepackage{tikz}

\usetikzlibrary{arrows, positioning, shapes.geometric, calc}

\sethlcolor{yellow!30}
\ADsetup{
title    = {Implementation and Analysis of Different Geomagnetic Field Models for Attitude Determination and Control System (ADCS) of a Satellite},
author   = {Hoor Bano$^{1}$ \cor{}, Tatiana Podladchikova$^{1}$, Bisma Sajid$^{1}$, and Dmitry Ris$^{1}$}, 
address  = {
  1. Skolkovo Institute of Science and Technology, Bolshoy Boulevard 30, bld. 1, 121205 Moscow, Russia \sep },
email    = {Hoor.Bano@skoltech.ru},
abstract = {
An Attitude Determination and Control System is essential for orientation stability and performance of slew maneuvers on the satellite. This research focuses on comparing two different geomagnetic field models, Direct Dipole Model and International Geomagnetic Reference Field Model, for modeling of magnetometer and magnetorquers. Both these magnetic field models are compared and analyzed for two satellite attitude cases: orientation stability and unloading of reaction wheels. Magnetometer modeling is utilized to get sensor data for attitude determination and control to attain orientation stability. Whereas, the magnetorquer model aids in reaction wheel unloading, by performing the required actuation on the satellite, upon interaction with the Earth’s magnetic field. The study offers a comprehensive lookout on the impact of geomagnetic field models on the overall ADCS performance, incorporating both attitude estimation and control via the sensor and actuator modeling. Apart from this, valuable insights are gained into selecting optimal models based on specific mission requirements and available computational resources. Finally, this comparison and analysis results in unique findings for an actual future satellite mission, that is to be launched soon.
  },
keywords = {
  Direct Dipole Model \sep Geomagnetic field model \sep IGRF \sep Magnetometer \sep Magnetorquer \sep Orientation stability \sep Unloading algorithm },
}
 
\begin{document}

\maketitle  


\Nomenclature

\begin{center}
\begin{tabular}{|P{.45\linewidth}|P{.45\linewidth}|}
\hline
$\mathbf{J}$ & Moment of Inertia ($kg.m^2$) \\
\hline
$\mathbf{M}$ & Torque ($Nm$)\\
\hline
$q_0$ & Scalar Quaternion Component\\
\hline
$q_1, q_2, q_3$ & Vector Quaternion Components\\
\hline
$\boldsymbol{\omega}$ & Satellite Angular Velocity ($deg/sec$) \\
\hline
$\mathbf{B_{mtm}}$ & Magnetometer Model ($T$)\\
\hline
$\mathbf{r_{ss}}$ & Sun sensor Model \\
\hline
$\boldsymbol{x}$ & State Vector\\
\hline
$\boldsymbol{\tilde{x}}$ & Reduced State Vector\\
\hline
\end{tabular}
\end{center}

\section{Introduction}

The design, analysis and integration of an accurate geomagnetic field model, in a satellite Attitude Determination and Control System (ADCS) design, is critical to its adequate performance and accuracy. To be as close to the actual environmental conditions that the satellite would be exposed to in space, it is necessary to properly formulate the geomagnetic field models, all the while catering for the computational limits. Various geomagnetic field models, such as the International Geomagnetic Reference Field (IGRF), World Magnetic Model (WMM) and Enhanced Magnetic Model (EMM) are quite accurate in modeling the Earth's magnetic field. Due to power constraints of smaller satellites like CubeSats, simpler low-computational dipole models like the Tilted dipole and Direct Dipole models, are also commonly used in satellite ADCS to accurately estimate the Earth's magnetic field for orientation purposes. 

The ADCS, being a core satellite system, aids in a number of functions like station keeping and performing slew-maneuvers, to name a few. Various ADCS modes require geomagnetic field modeling for attitude determination, and in some cases, to unload the actuators like reaction wheels. An ADCS essentially relies on a combination of sensors and actuators to achieve its objectives. Among these components, magnetometers and magnetorquers have become vital elements of modern satellite systems in Low Earth Orbits, both of them utilizing Earth's magnetic field to operate. A magnetometer provides satellite's orientation data based on the Earth's magnetic field measurement. On the other hand, magnetorquers produce the required torque to adjust the satellite's attitude upon interaction with Earth's magnetic field. 

Earth's magnetic field is generated by the outer core fluid in a self-exciting dynamo process. Additionally, the sources of Earth's magnetic field also include its crust, ionosphere and magnetosphere \cite{bgs_magnetic_field}. Antipov and Tikhonov \cite{Antipov2013} were able to derive concise and practical mathematical formulations that allow the multi-pole tensor components to accurately represent the gradient tensor components and the geomagnetic induction vector to any desired degree. Reference \cite{luo2023attitude} presents a methodology for satellite attitude control by using the first-order approximate IGRF model to calculate the geomagnetic field vector, in the event of magnetometer failure and computational constraints.

Numerous studies have emerged in the past comparing different geomagnetic field models for the general application of satellite attitude control. Navabi and Barati \cite{navabi2017mathematical} implemented and compared different models, such as Simplified Dipole, Centered Dipole, Dipole, Quadrupole, Octupole, Sedecimupole and other Reference Models. A comparison study was also conducted to show effects of orbit inclination, altitude, latitude and longitude on the geomagnetic field models. In conclusion, higher order models proved to be more accurate for satellite attitude control. However, the research also deduces that the choice of the model is a trade-off between accuracy and computational effort, and must be chosen based on mission scenarios. 

Ovchinnikov et al. \cite{Ovchinnikov2018} also provides a very informative comparison between four different magnetic field models of the Earth for satellite angular motion research. Conclusions are drawn and articulated that the usage of these different models depends on different cases of satellite attitude problems such as nature of the motion, analysis method and reference frames. Reference \cite{ivanov2016advanced} shows fully controlled magnetic system comprising of magnetometer and magnetorquers. Inclined dipole model is used for geomagnetic field modeling, and achieves three axis control with an accuracy of 2$^{\circ}$-10$^{\circ}$. 

Research presented in \cite{wisniewski_satellite} provides control laws for three-axis stabilization by magnetic actuation. $10^{th}$ order spherical harmonic model has been used to calculate the geomagnetic field from \cite{wertz1990spacecraft}. 
A constant gain and a time-varying linear controller are designed using the observation that the geomagnetic field on near polar orbit is roughly periodic. 

Unloading of the reaction wheels by momentum dumping is essential for their optimal and efficient performance. Markley and Crassidis \cite{MarkleyCrassidis2014} discuss this process, using magnetorquers. They also mention the requirements for using magnetic control to ensure three-axis control of the satellite. These highlight the importance of avoiding activation of magnetorquers near the geomagnetic poles or the geomagnetic equatorial plane, where control effectiveness is reduced. 

Acar and Horri \cite{acar2013optimal} perform momentum dumping for reaction wheels by employing controllers that balance rapid unloading with energy efficiency, using magnetic torquers, thrusters, or a combination of both. Meanwhile in \cite{chujo2024integrated}, a propellant-free momentum unloading strategy is implemented by combining a solar sail with a single-axis gimbal mechanism, which shifts the attitude equilibrium to minimize reaction wheel momentum buildup.

Review of the previous studies indicates that even though the comparison of the geomagnetic field models has been studied quite a few times, the comparative analysis covering a complete ADCS model, with accompanying actuators, sensors and especially the actuator de-saturation has been rarely touched upon. An important factor when deciding on the geomagnetic field model, for a particular satellite mission, is its impact on the overall ADCS performance. The key challenge can be selecting a model that not only provides efficient estimation and control, but also keeps into account the computational resources available.

In this study we aim to overcome the challenges of selecting an optimal geomagnetic field model and consider the following tasks. First, we perform comprehensive comparison of geomagnetic field models in ADCS. While previous studies have compared geomagnetic field models, this research uniquely integrates both sensor modeling (magnetometers) and actuator control (magnetorquers for reaction wheel unloading) within a complete ADCS framework. Second, we analyze reaction wheels unloading algorithm with varying magnetic field models. The study evaluates how different geomagnetic models impact the unloading of reaction wheels, optimizing control efficiency by determining the best unloading intervals. Third, we study the effect on overall ADCS performance. Unlike prior work that focused primarily on attitude estimation or control separately, this study examines the impact of varying magnetic field models on both attitude stabilization and actuator performance, incorporating Extended Kalman Filter (EKF) for enhanced estimation accuracy. And finally, the study’s implementation on the Skoltech-F CubeSat mission demonstrates its real-world applicability by offering valuable insights into selecting an optimal geomagnetic model that balances accuracy with computational efficiency, which is a critical trade-off for small satellites with limited resources.

The current research focuses on implementing and analyzing two different geomagnetic field models, i.e. Direct Dipole model and International Geomagnetic Reference Field (IGRF) Model, for the cases of attitude stabilization and unloading of reaction wheels. A comparative study is drawn between the two geomagnetic field models, to determine the better model in terms of accuracy, stability and mission requirements. In case of attitude stabilization, a three-axis magnetometer is used as a sensor for the attitude determination, prior to implementing the reaction wheel-based control. While in order to de-saturate the reaction wheels, an array of three magnetorquers is used with the B-dot controller to produce the required unloading torque. Both these tasks require geomagnetic field modeling.

Section \ref{Sec2} of the paper includes a detailed description of the entire satellite system which includes the mathematical models of satellite rotational dynamics, incorporating sensors and actuators, the geomagnetic field models, the control laws and the estimation algorithm. This is followed by simulation of the developed models and discussion of the obtained results in Section \ref{Sec3}, and finally the conclusion is drawn in Section \ref{Sec4}.

\section{Mathematical Modeling}
\label{Sec2}

The ADCS structure followed in this research is based on the work conducted in  \cite{bano2024development}, where the control goal was to align the orbital frame with the satellite body frame, with an angular accuracy of 1 degree. Magnetometer and sun sensors are used as sensors for attitude determination. An Extended Kalman Filter is used for attitude estimation. For attitude control, a Proportional-Derivative (PD) and B-dot control laws are used, with an array of four reaction wheels in a tetrahedron configuration for stabilization and 3 magnetorquers in an orthogonal configuration for reaction wheel unloading. Given below in Figure \ref{Fig1} is a framework diagram for the ADCS structure followed in this paper.

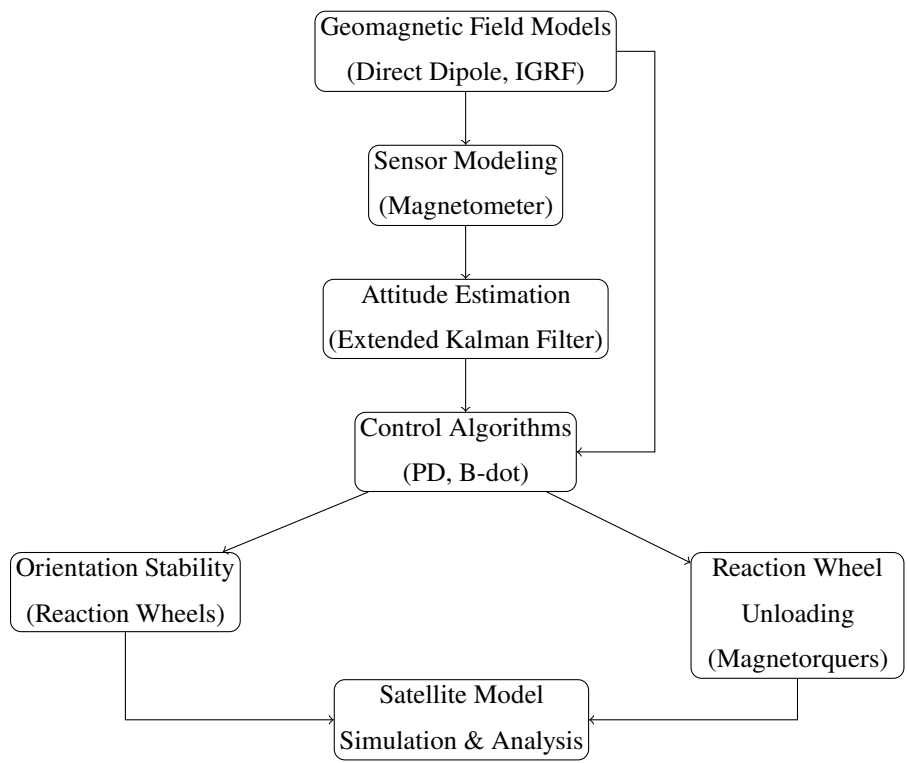
\begin{figure}[hbt!]
\begin{center}
\begin{tikzpicture}[
  node distance=0.7cm,   
  auto,
  every node/.style={
    draw,
    rectangle,
    rounded corners,
    align=center,
    font=\small,        
    inner sep=2pt       
  }
]

  \node (geomag)
    {Geomagnetic Field Models\\[-2pt](Direct Dipole, IGRF)};

  \node[below=of geomag] (sensor)
    {Sensor Modeling\\[-2pt](Magnetometer)};

  \node[below=of sensor] (estimation)
    {Attitude Estimation\\[-2pt](Extended Kalman Filter)};

  \node[below=of estimation] (control)
    {Control Algorithms\\[-2pt](PD, B-dot)};

  \node[below left=of control, xshift=-1.0cm, yshift=-0.3cm,
        minimum width=2.8cm] (stability)
    {Orientation Stability\\[-2pt](Reaction Wheels)};

  \node[below right=of control, xshift=1.0cm, yshift=-0.3cm,
        minimum width=2.8cm] (unloading)
    {Reaction Wheel\\[-2pt]Unloading\\[-2pt](Magnetorquers)};

  \node[below=of $(stability)!0.5!(unloading)$, yshift=-0.3cm] (simulation)
    {Satellite Model\\[-2pt]Simulation \& Analysis};


  \draw[->] (geomag) -- (sensor);
  \draw[->] (sensor) -- (estimation);
  \draw[->] (estimation) -- (control);

  \draw[->] (geomag.east) -- ++(0.5cm,0) coordinate (A)
       -- (A |- control.east) -- (control.east);

  \draw[->] (control) -- (stability);
  \draw[->] (control) -- (unloading);

  \draw[->] (stability.south) |- (simulation.west);
  \draw[->] (unloading.south) |- (simulation.east);

\end{tikzpicture}
\caption{Framework diagram illustrating the structure of the ADCS system.}
\label{Fig1}
\end{center}
\end{figure}

\subsection{Reference Frames}
\subsubsection*{Earth Centered Inertial (ECI) Frame}
This inertial frame moves with Earth’s center, but does not rotate. In ECI frame, x-axis points in the vernal equinox direction, z-axis is along the direction of the Earth’s North Pole, and y-axis completes the right-handed system \cite{yang2019spacecraft}. 

\subsubsection*{Body Frame}
Body frame rotates with the satellite itself, and is fixed with the satellite body, with its origin at the center of mass of the satellite. 

\subsubsection*{Orbital Frame}
This reference frame represents the position of the satellite in an orbit relative to a specific reference point. The x-axis coincides with the velocity vector of the circular orbit, z-axis points away from the center of mass of the Earth (or radial vector) and y-axis completes the right-handed system.

\subsection{Geomagnetic Field Models}

Geomagnetic field models are mathematical representations of Earth's magnetic field, which is generated by the motion of molten iron in the outer core. These models provide a detailed description of the magnetic field at different locations on and around the Earth, especially for satellite systems in Low Earth Orbits (LEO). Two different geomagnetic field models are used here; Direct Dipole model and International Geomagnetic Reference Field Model (IGRF).

\subsubsection{Direct Dipole Model}

Direct Dipole model is one of the simplest models, providing first order approximation of the Earth’s magnetic field. It assumes the Earth to be a tilted magnetic dipole. Geomagnetic induction vector is given as \cite{Ovchinnikov2018}:

\begin{equation}
\mathbf{B} = -\frac{\mu_{e} \mu_{0}}{4\pi |\mathbf{r_{sat}}|^5}
(\mathbf{k}|\mathbf{r_{sat}}|^2 - 3(\mathbf{k}\mathbf{r_{sat}})\mathbf{r_{sat}})
\label{1}
\end{equation}

Here, $\mu_{e} = 7.94 \times 10^{22} Am^{2}$ is the magnetic dipole moment of the Earth, $\mu_{0} = 1.257 \times 10^{-6} NA^{-1}$ is the vacuum permeability, $\mathbf{r_{sat}}$ is the orbital radius of the satellite, and $\mathbf{k}$ is the Earth's dipole vector. 

For satellite attitude control, the Direct Dipole model offers an adequate estimate for the geomagnetic field, and hence, for satellite orientation states. The absence of Earth’s rotation and the orbital precession are also assumed for this model. The geomagnetic induction vector in orbital frame is given as \cite{navabi2017mathematical}:

\begin{equation}
\mathbf{B^{orb}_{model}} = \frac{\mu_{e} \mu_{0}}{4\pi |\mathbf{r_{sat}}|^3}
\begin{bmatrix}
\cos(u) \sin(i) \\
\cos(i) \\
-2\sin(u) \sin(i)
\end{bmatrix}
\label{2}
\end{equation}

Here, $u$ is the argument of latitude of the orbit and $i$ is the orbit inclination.

\begin{figure}[hbt!]
\centerline{\includegraphics[width=20pc]{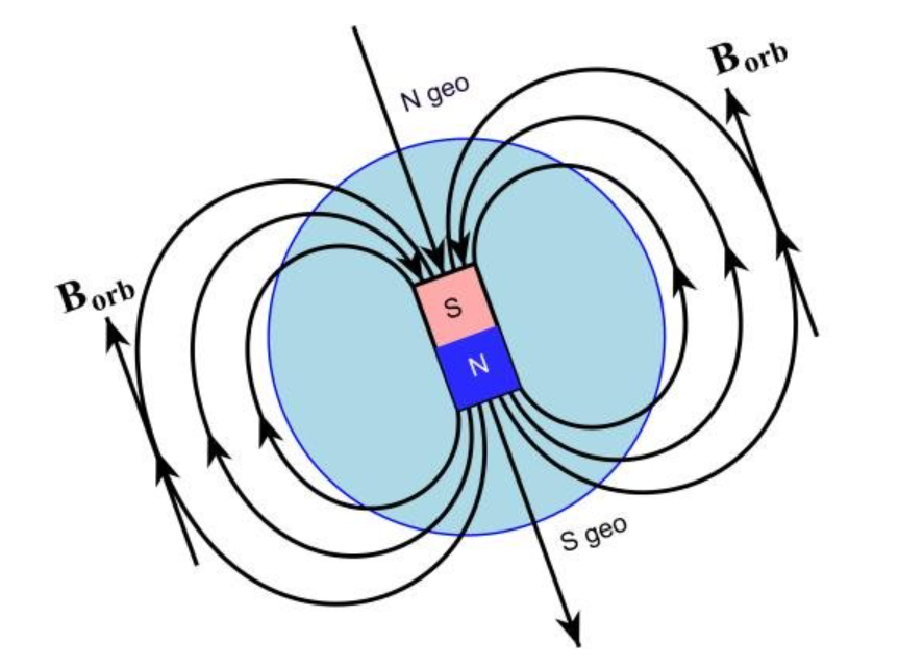}}
\caption{Simple Representation of Earth's Magnetic Field as a Dipole \cite{doroshin2013exact}}
\label{Fig2}
\end{figure}

Figure \ref{Fig2} shows representation of the Earth's magnetic field as assumed in the Direct Dipole model. 

\subsubsection{International Geomagnetic Reference Field (IGRF) Model}

The Earth's magnetic field is composed of several magnetic fields generated by a variety of sources, including the Earths fluid core, its crust and the upper mantle, the ionosphere and the magnetosphere. Hence, it is essential to develop and simulate this accurate magnetic field model. The International Association of Geomagnetism and Aeronomy (IAGA) has provided and regularly updates the International Geomagnetic Reference Field (IGRF) Model, which the latest being IGRF-13.

The potential function $V(r, \theta, \phi, t)$ is represented as a finite series expansion in terms of spherical harmonic coefficients, $g(t)^m_n$ and $h(t)^m_n$ also known as the Gauss coefficients. It is given as \cite{alken2021international}: 
\begin{equation*} 
V(r, \theta, \phi, t)  = R_{Earth} \sum\limits^N_{n=1} \sum\limits^n_{m=0} (\frac{R_{Earth}}{r})^{n+1} \; 
\end{equation*} 

\begin{equation} \label{3}
P^m_n \; cos(\theta) \;
(g(t)^m_n \; cos(m\phi) + h(t)^m_n \; sin(m\phi)) 
\end{equation}

Here, r, $ \theta$, $\phi$ refer to coordinates in a geocentric spherical coordinate system, with r being radial distance from the center of the Earth, and $ \theta$, $\phi$ representing geocentric co-latitude and longitude, respectively. $R_{Earth}$ = 6371.2 km is the mean radius of the Earth. The term $P^m_n cos(\theta)$ corresponds to the Schmidt semi-normalized associated Legendre functions of degree n and order m \cite{winch2005geomagnetism}. The parameter N specifies the maximum
spherical harmonic degree of expansion, and is chosen to be 13 for IGRF-13 model to account for small scale internal signals. 

\begin{figure}[hbt!]
\centerline{
\includegraphics[bb=0 0 5972 3742, width=20pc]{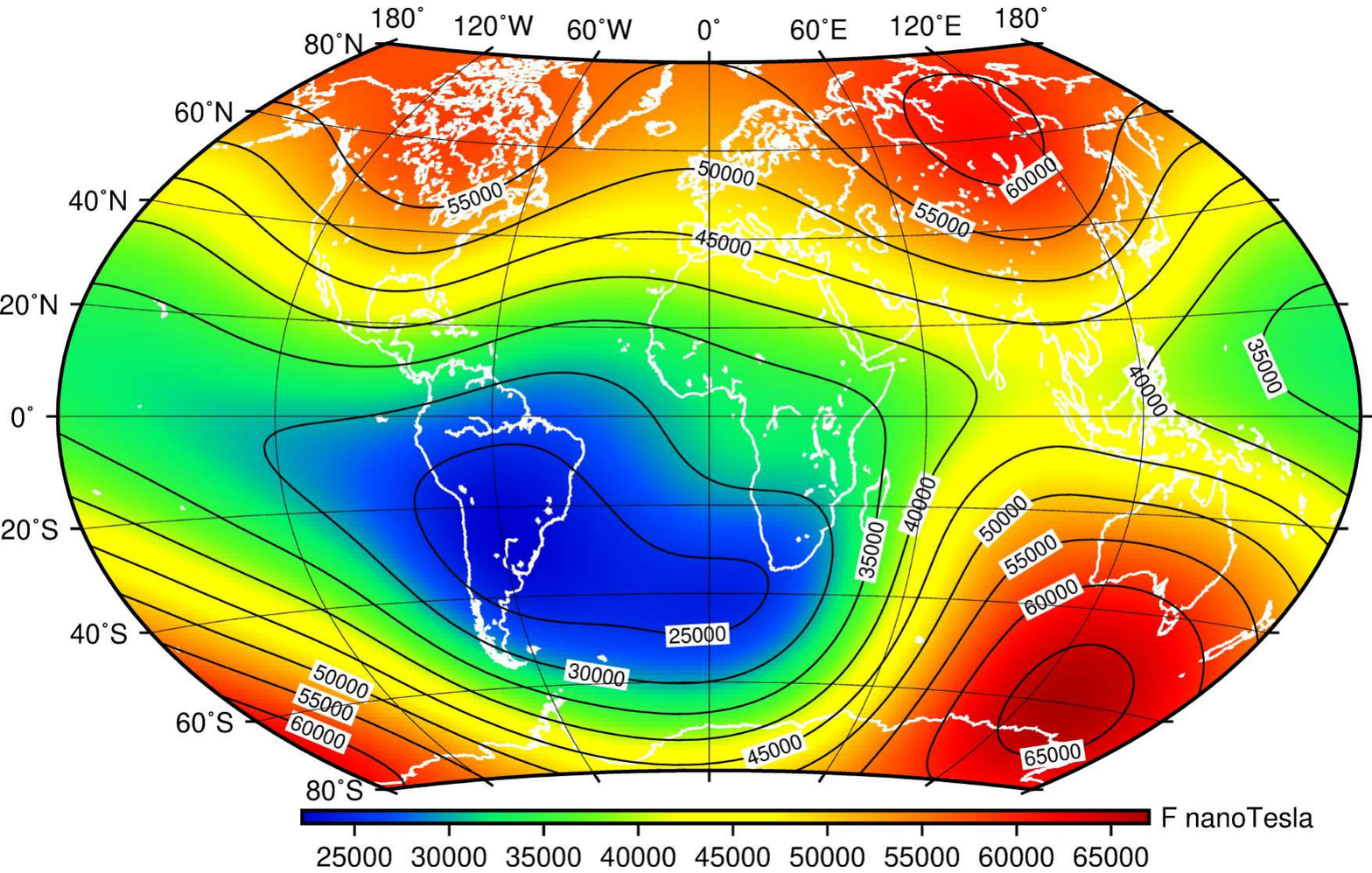}
}
\caption{Total Geomagnetic Field Intensity (nT) by IGRF Model \cite{igrf}}
\label{Fig3}
\end{figure}

Figure \ref{Fig3} represents an accurate model for Earth's geomagnetic field intensity as assumed by the IGRF model. 

\subsection{Satellite Attitude Model}
To implement different geomagnetic field models for satellite orientation control, we must first model the satellite attitude dynamics and kinematics. 
\subsubsection{Satellite Dynamics}

Based on rigid body dynamics, non-linear satellite dynamics equation is given as \cite{kluever2018space}:
\begin{equation}
    \mathbf{J}\boldsymbol{\dot{\omega}} = -\boldsymbol{\omega} \times (\mathbf{J}\boldsymbol{\omega} + \mathbf{h}) + \mathbf{M}
    \label{4}
\end{equation}

Here, $\mathbf{M}$ denotes the disturbance, control and unloading torque, $\mathbf{J}$ denotes the inertia tensor, $\mathbf{h}$  denotes the angular momentum when using reaction wheels and $\boldsymbol{\omega}$ denotes the angular rate of the satellite in the body frame.

\subsubsection{Satellite Kinematics}
A quaternion represents a rotation by a rotational angle around a rotational axis. It comprises four elements. 

\begin{equation}
\boldsymbol{q} = a + b\mathbf{i} + c\mathbf{j} + d\mathbf{k} 
\label{5}
\end{equation}

Here, the coefficients a, b, c and d are real numbers, whereas 1, $\mathbf{i}$, $\mathbf{j}$ and $\mathbf{k}$ are the basis vectors or basis elements. Quaternion provides a very efficient way of transformation to and from the body frame. Compared to Euler angles, it avoids singularity issues which can cause loss of degree of freedom in some axes. Quaternion kinematics is: 

\begin{equation}
\boldsymbol{\dot{q}} = \frac{1}{2}\,\boldsymbol{q} \circ {\boldsymbol{\omega}}
\label{6}
\end{equation}
$o$ represents the quaternion multiplication. The transformation of the satellite angular rate from the body frame to the inertial frame is given as \cite{yang2019spacecraft}:

\begin{equation}
\begin{bmatrix}
\dot{q}_0 \\
\dot{q}_1 \\
\dot{q}_2 \\
\dot{q}_3
\end{bmatrix}
= 0.5
\begin{bmatrix}
{q}_0 & -{q}_1 & -{q}_2 & -{q}_3 \\
{q}_1 & {q}_0 & -{q}_3 & {q}_2 \\
{q}_2 & {q}_3 & {q}_0 & -{q}_1 \\
{q}_3 & -{q}_2 & {q}_1 & {q}_0
\end{bmatrix}
\begin{bmatrix}
0 \\
\omega_1 \\
\omega_2 \\
\omega_3
\end{bmatrix}
\label{7}
\end{equation}

Here,  ${q}_0$ (the scalar component) and ${q}_1$, ${q}_2$ and ${q}_3$ (the vector components) represent an ordered quadruple of real numbers a, b, c and d. Eqn. \ref{7} represents the satellite kinematics equation for frame transformation. 

\subsection{Sensor Modeling}
Sensor models enable us to model the measurements from different sensors for attitude determination. For this research, we model a 3-axis magnetometer and an array of five sun sensors. 

\subsubsection{Magnetometer Modeling}

The environment model for Earth’s magnetic field is generated by using either the Direct Dipole Model from Eqn. \ref{2} or IGRF model from Eqn. \ref{3}, with incorporation of the environmental noise. It is given with respect to the satellite body frame as:

\begin{equation} \label{8}
\mathbf{B_{env}} = \mathbf{{B}^{b}_{model}} + \mathbf{N_{env}}(\mathrm{\mu_{env}},\mathrm{\sigma_{env}})
\end{equation}

Here, $\mathbf{N_{env}}$ denotes the normally distributed vector for environmental noise with mean $\mathrm{\mu_{env}}$ and standard deviation $\mathrm{\sigma_{env}}$. 

The sensor model for the magnetometer in the satellite body frame uses the environmental model from Eqn. \ref{8} with added inherent noise of the sensor itself. It is given as: 

\begin{equation} \label{9}
\mathbf{B_{mtm}} = \mathbf{{B}_{env}} + \mathbf{N_{mtm}}(\mathrm{\mu_{mtm}}, \mathrm{\sigma_{mtm}})
\end{equation}

Here, $\mathbf{N_{mtm}}$ denotes the normally distributed vector for sensor noise with mean $\mathrm{\mu_{mtm}}$ and standard deviation $\mathrm{\sigma_{mtm}}$.

\subsubsection{Sun Sensor Modeling}

A Sun sensor is a navigational instrument used by a satellite to detect the position of the Sun for orientation determination. To model the sun sensors, the Sun direction vector and eclipse prediction must be modeled. Vallado \cite{vallado2013fundamentals} has proposed a model to calculate the Sun direction vector by calculating the ecliptic longitude and obliquity between the Sun and Earth's ecliptic frame in ECI frame. He also used line of sight analysis to determine whether the satellite is in the sunlit or eclipse part of the orbit.

For sun sensor measurements, the Sun's presence within the field of view of a specific sensor is verified by calculating the incidence angle as described in \cite{bano2024development}. If valid, the Sun direction vector is adjusted for sensor noise using a rotation operator. 

\begin{equation} \label{10}
   \mathbf{r_{ss}} = \mathrm{R(\mathrm{\theta_{ss}})}\mathbf{r_{model}}
\end{equation}

$\mathbf{r_{ss}}$ is the Sun direction vector rotated by sensor noise in the satellite body frame, derived from the modeled unit Sun direction vector $\mathbf{r_{model}}$ via the rotation operator $\mathrm{R(\theta_{ss}})$. Here, $\mathrm{\theta_{ss}}$ is a random rotation angle with mean 0 and standard deviation $\mathbf{\sigma_{ss}}$. The nonlinear dependence in Eqn. \ref{10} is addressed by validating sun sensor data before processing.

\subsection{Control Algorithms} \label{Sec2.5}
To get a complete model of the satellite dynamics, we must determine the control, disturbance and unloading torque in Eqn. \ref{4}. For this purpose, Proportional-Derivative (PD) Control and B-dot Control are used. 
\subsubsection{Proportional-Derivative (PD) Control}

The PD control gains can be calculated from the Linear Quadratic Regular (LQR), an optimal feedback controller, which minimizes the quadratic cost function. Estimated states from the EKF are used in the PD algorithm to calculate the required control torque. It is given as \cite{ivanov2012testing}: 

\begin{equation} \label{11}
\mathbf{M_{ctrl}} = -\mathbf{M_{ext}} - k_{\omega}J\boldsymbol{\omega_{err}} - k_qJ\boldsymbol{q_{err}}
\end{equation}

$\boldsymbol{\omega_{err}}$ and $\boldsymbol{q_{err}}$ denote the error (difference) calculated between the estimated and the required angular rates and quaternions, respectively, that has to be reduced by the controller. $k_{\omega}$ and $k_q$ denote the derivative and proportional control gains, respectively. $\mathbf{M_{ext}}$ is the external disturbance torque that the controller has to compensate.

\subsubsection{B-dot Control}

To avoid wheel saturation over time, unloading algorithm for reaction wheels angular momentum must be implemented. For this purpose, magnetorquers will be used by a B-dot controller to generate the required magnetic moment, and hence, the unloading torque.

B-Dot controller is a simple and widely used control algorithm to detumble satellites as well as manage the reaction wheel momentum. This controller computes the magnitude and direction of current to be passed through magnetorquers to generate the moment that will interact with the Earth’s magnetic field \cite{sharma2021simulation}. The moment required for reaction wheel angular momentum dumping is given as:

\begin{equation} \label{12}
\mathbf{m_{req}} = \mathbf{h} \times \mathbf{B_{mtm}}
\end{equation}

Here, $\mathbf{h}$ denotes the angular momentum of the reaction wheels and $\mathbf{B_{mtm}}$ denotes the sensed magnetic field from the magnetometer.

The maximum possible ratio of the moment is given as:

\begin{equation} \label{13}
\mathrm{Max \: Ratio} = max\left(\frac{|\mathbf{m_{req}|}}{m_{max}}\right)
\end{equation}

Here, the maximum moment $m_{max}$ that can be generated by the magnetorquers is based on the voltage available. It is explained and given in magnetorquers modeling in Section \ref{Sec2.6}.

The actual moment produced is given as:

\begin{equation} \label{14}
\mathbf{m_{actual}} = \frac{\mathbf{m_{req}}}{\mathrm{Max \: Ratio}}
\end{equation}

The unloading torque produced by the moment for de-saturation of wheels is given as:
\begin{equation} \label{15}
\mathbf{M_{unload}} = \mathbf{m_{actual}} \times \mathbf{B_{env}}
\end{equation}

\subsection{Actuator Modeling} \label{Sec2.6}
Actuator modeling is essential to allow the satellite to perform actuation based on the torque calculated by the control algorithms in Section \ref{Sec2.5}.  
\subsubsection{Reaction Wheels} 

Reaction wheels work on the principal of conservation of angular momentum. The torques of reaction wheel are generated from acceleration or deceleration of the rotational flywheel \cite{yang2019spacecraft}. The required torque from the control algorithm in Eqn. \ref{11} is sent to the reaction wheels for actuation. As reaction wheels have a maximum torque $M_{max}$  and angular momentum constraint $h_{max}$, the torque produced by each reaction wheel $\mathbf{M_{rw}}$ is based upon the limits of these specifications, described in \cite{bano2024development}. 

$\mathbf{M_{allocated}}$ is the torque allocated for each reaction wheel in a specific axis based on the chosen array configuration. Four reaction wheels are arranged in a tetrahedron configuration for research. This configuration proves to be the most efficient as it consumes the least amount of power and provides most actuator failure robustness in terms of system performance \cite{kok2012comparison}.

The torque $\mathbf{M_{allocated}}$ in an array is given as:  
\begin{equation} \label{16}
\mathbf{M_{allocated}} = \frac{A^{-1}\:W^{-1}}{W\:A^{-1}\:W^{-1}}\: \mathbf{M_{ctrl}}
\end{equation}

Here, $A$ is the matrix of the torque quadratic form to be minimized at torque allocation, $W$ denotes the installation matrix and $\mathbf{M_{ctrl}}$ is the required control torque. For tetrahedron configuration, $W$ is given as:

\begin{equation} \label{17}
W =                                 
\begin{bmatrix}
\sqrt{\frac{8}{9}}  & -\sqrt{\frac{2}{9}} & -\sqrt{\frac{2}{9}} & 0 \\
0 & \sqrt{\frac{2}{3}} & -\sqrt{\frac{2}{3}} & 0 \\
-\frac{1}{3} & -\frac{1}{3} & -\frac{1}{3} & 1 
\end{bmatrix}
\end{equation}

\subsubsection{Magnetorquers}
Magnetorquers are generally planar coils of uniform wire rigidly placed along the satellite body axes. When electricity passes through the coils, a magnetic dipole is created \cite{yang2019spacecraft}. One of the main uses of magnetorquers is to dump excess momentum induced by external disturbances. Their operation is based on the interaction between the magnetic field generated by the coils and the magnetic field of Earth. The maximum moment $m_{max}$ that can be generated by the magnetorquers is based on the maximum voltage available.

The maximum moment $m_{max}$  is given as:

\begin{equation} \label{18}
m_{max} = \frac{V_{max}}{\mathrm{Res}} (n\,A_{coil}) \: A_{mtq}
\end{equation}

Here, $V_{max}$ is the maximum voltage available from the Electrical Power System (EPS), $\mathrm{Res}$ is the resistance of the wire, $n$ is the number of turns of the wire, $A_{coil}$ is the  vector area of the coil and $A_{mtq}$ denotes the axis of the magnetorquer around which it produces the moment in the body frame. It is also to be noted that magnetorquers are also used in an array, generally with an orthogonal configuration.

\subsection{Extended Kalman Filter}

An attitude estimation algorithm must be used to bridge the gap between attitude determination and attitude control. For this purpose, an Extended Kalman Filter is used. The dynamic system model is used to estimate the state vector $\boldsymbol{x(t)}$ based on noisy measurements $\mathbf{z(t)}$, given as: 
\begin{equation}\label{19}
\dot{\boldsymbol{x}} = \mathbf{f(x,t)} + \mathbf{G}\mathbf{w(t)} 
\end{equation}

$\mathbf{f(x, t)}$ is given as \cite{yang2019spacecraft}:

\begin{equation} \label{20}
 \mathbf{f(x, t)} = 
\begin{bmatrix}
\boldsymbol{\dot{q}} \\
\dot{\boldsymbol{\omega}}
\end{bmatrix}
=
\begin{bmatrix} 
\frac{1}{2}\,\boldsymbol{q}\circ{\boldsymbol{\omega}} \\
   \mathbf{J}^{-1} (-\boldsymbol{\omega} \times (\mathbf{J}\boldsymbol{\omega} + \mathbf{h}) + \mathbf{M})
\end{bmatrix}
\end{equation}

The input matrix $\mathbf{G}$ is given as:
\begin{equation} \label{21}
{\mathbf{G}} = \begin{bmatrix}
\mathbf{0_{3\times3}} \\
\mathbf{J}^{-1}
\end{bmatrix}
\end{equation}

$\mathbf{w(t)}$ is the Gaussian random process with zero mean and covariance matrix $\mathbf{D}$, which describes the random nature of the process evolution. The measurement $\mathbf{z(t)}$ and the true state vector $\boldsymbol{x(t)}$ are related by the measurement equation. It is given as:
\begin{equation}\label{22}
\mathbf{{z}(t)} = \mathbf{h(x,t)} + \mathbf{v(t)}
\end{equation}

Here, vector function $\mathbf{h(x,t)}$ links the true state vector to the measurement vector, and $\mathbf{v(t)}$ represents Gaussian measurement noise with covariance matrix $\mathbf{R}$.

For the process model, dynamics from Eqn. \ref{4} and kinematics from Eqn. \ref{7} are used. For the observations, magnetometer and sun sensor measurements from Eqn. \ref{9} and Eqn. \ref{10} are used, respectively. 

The state vector (7-dimensional), for calculation of the predicted states and for the observation vector (in correction step), comprises the four quaternion and three angular velocity components, given as:
\begin{equation}\label{23}
\boldsymbol{x} = \begin{bmatrix}
\boldsymbol{q} \\
\boldsymbol{\omega}
\end{bmatrix}
\end{equation}

However, the reduced state vector $\boldsymbol{\tilde{x}}$ (6-dimensional) for the evolution matrix and corrected states comprises the three vector components of the quaternion and three angular velocity vectors. This is because a quaternion follows the normalization constraint and the fourth scalar component can be calculated from it, given as:

\begin{equation}\label{24}
q_{0} = \sqrt{1 - q_1^{2} - q_2^{2} - q_3^{2}}
\end{equation}

TRIAD (Tri-axial Determination) is a deterministic model that uses two vector measurements to determine the orientation. In our case, these measurements comprise the magnetometer and sun sensor readings. Gokcay and Hajiyev \cite{gokcay2022triad} gave a simple algorithm to determine the quaternion from sun sensor and magnetometer measurements, which will be used to initialize the filter.

\subsubsection{Prediction}

The predicted state $\boldsymbol{{x_{k,k-1}}}$ is calculated by integration of the non-linear process model from Eqn. \ref{20}, given as:

\begin{equation}\label{25} 
\boldsymbol{{x_{k,k-1}}} = \int_{k-1}^{k} \mathbf{f(\boldsymbol{x_{k-1,k-1}}, t)}\:dt
\end{equation}

Here, the 7-dimensional vector  $\boldsymbol{x_{k-1,k-1}}$ represents the updated state estimate, after incorporating measurements at time step $k-1$ and the derivation of the fourth component of the quarternion. 

Prediction of the error covariance matrix  $\mathbf{P_{k,k-1}}$ is given as \cite{Ovchinnikov2014}:

\begin{equation} \label{26}
\mathbf{P_{k,k-1}} = \boldsymbol{\Phi_{k}} \mathbf{P_{k-1,k-1}} \boldsymbol{\Phi_{k}}^T + \mathbf{Q_{k}}
\end{equation}

Here, $\mathbf{P_{k-1,k-1}}$ represents the covariance matrix of the estimation error for the 6-dimensional vector $\boldsymbol{\tilde{x}_{k-1,k-1}}$, updated after incorporating measurements at time step $k-1$ and $\mathbf{Q_{k}}$ denotes the process noise covariance matrix. The initial value $\mathbf{P_{o}}$ for the error covariance matrix represents the initial uncertainty in the states $\boldsymbol{\tilde{x}_o}$. The state transition matrix $\boldsymbol{\Phi_{k}}$ relates the state vector at the previous step to the next one. It is defined as the result of discretizing a linear continuous system with the dynamic matrix $\mathbf{F_{k}}$, which in turn is obtained by linearizing the satellite model, given in \cite{Ovchinnikov2014}:
\begin{equation}\label{27}
\boldsymbol{\Phi_{k}} = \mathbf{I_{6\times6}} + \mathbf{F_{k}} \Delta t
\end{equation}

Here, $\Delta t$  is the time interval of each iteration. The matrix $\mathbf{F_{k}}$ characterizes the rate of change of the original nonlinear state function evaluated at the 6-dimensional state  estimate vector $\boldsymbol{\tilde{x}}$ at time $t_k$. It is given as \cite{ivanov2012testing}: 
\begin{equation} \label{28}
    \mathbf{F_{k}} = \begin{bmatrix}
        -\mathbf{W_{\omega}} & \frac{1}{2} \mathbf{I_{3\times3}} \\  - k_{q}\mathbf{I_{3\times3}} & -k_{\omega}\mathbf{I_{3\times3}}
    \end{bmatrix}
\end{equation}

Here, $k_{\omega}$ and $k_q$ denote the derivative and proportional control gains, respectively.  $\mathbf{W_{\omega} }$  represents the skew-symmetric matrix for $\boldsymbol{\omega}$, given as:

\begin{equation} \label{29}
    \mathbf{W_{\omega}} = \begin{bmatrix}
        0 & -\omega_3 & \omega_2 \\ \omega_3 & 0 & -\omega_1 \\ -\omega_2 & \omega_1 & 0
    \end{bmatrix}
\end{equation}

In Eqn. \ref{26}, the covariance matrix for process noise $\mathbf{Q_{k}}$ is given as \cite{mikhaylov2014autonomous}:

\begin{equation} \label{30}
    \mathbf{Q_{k}} = \boldsymbol{\Gamma_{k}} \,\mathbf{D}\,\boldsymbol{\Gamma_{k}}^{T}
\end{equation}

$\boldsymbol{\Gamma_{k}}$ represents the process noise transition matrix, given as \cite{mikhaylov2014autonomous}:

\begin{equation} \label{31}
\boldsymbol{\Gamma_k} = \mathbf{G}\:\Delta t + \mathbf{F_{k}}\mathbf{G}\frac{\Delta t^{2}}{2}
\end{equation}

Here, $\mathbf{G}$ is the input matrix. $\mathbf{D}$ quantifies the uncertainty associated with the disturbances, given as:

\begin{equation} \label{32}
\mathbf{D} = \sigma^2_{dist}\mathbf{I_{3\times3}}
\end{equation}

$\mathrm{\sigma_{dist}}$ denotes the standard deviation of the environmental disturbance torques. 

\subsubsection{Correction}
The correction phase of the filter uses noisy sensor measurements to correct the predicted states. The Kalman gain $\mathbf{K_k}$ is given as: \cite{Ovchinnikov2014}: 

\begin{equation} \label{33}
\mathbf{
K_k = P_{k,k-1}H^{T}_k(H_kP_{k,k-1}H^{T}_k + R)^{-1}}
\end{equation}

Here, $\mathbf{P_{k,k-1}}$ is the predicted error covariance matrix, $\mathbf{H_k}$ is the observation matrix and $\mathbf{R}$ is the measurement noise covariance matrix.

The estimated states are given as:

\begin{equation} \label{34}
\boldsymbol{\tilde{x}_{k,k}} = \boldsymbol{\tilde{x}_{k,k-1}} + \mathbf{K_k(z_k-h(x_{k,k-1},t_k)})
\end{equation}

 Here, $\boldsymbol{\tilde{x}_{k,k-1}}$ denotes the predicted states, $\mathbf{z_k}$ are the observations and the vector $\mathbf{h(x_{k,k-1},t_k)}$ provides non-linear relation between the observations and the states.

Correction of the error covariance matrix $\mathbf{P_{k,k}}$ is given as: 

\begin{equation} \label{35}
\mathbf{P_{k,k} = (I_{6\times6} - K_kH_k)P_{k,k-1}}
\end{equation}

The observations $\mathbf{z_k}$ contain the sensor measurements from (9) and (10), given as: 
\begin{equation} \label{36}
\mathbf{z_k} = \begin{bmatrix}
{\mathbf{\hat{B}_{mtm}}}\\
\mathbf{r_{ss}}
\end{bmatrix}
\end{equation}

For the calculation of the vector $\mathbf{h(x_{k,k-1},t_k)}$, the models (magnetic field and sun direction) in orbital frame are converted to body frame, and the transformation is performed using Eqn. \ref{7} by the predicted quaternions from Eqn. \ref{25}. 
\begin{equation} \label{37}
\mathbf{h(x_{k,k-1},t_k)} = 
\begin{bmatrix}
\boldsymbol{q^{o\rightarrow b}}\circ\mathbf{\hat{B}^{orb}_{model}}\circ\boldsymbol{\tilde{q}^{o\rightarrow b}} \\
\boldsymbol{q^{o\rightarrow b}} \circ\mathbf{r_{sun}}\circ\boldsymbol{\tilde{q}^{o \rightarrow b}}
\end{bmatrix}
=
\begin{bmatrix}
\mathbf{\hat{B}^b_{model} }\\
\mathbf{r_{sun}^b}
\end{bmatrix}
\end{equation}

The observation matrix $\mathbf{H_k}$ is the Jacobian of the observation vector function $\mathbf{h(x_{k,k-1},t_k)}$ with respect to the state vector $\boldsymbol{\tilde{x}}$ at time $t_k$, given as \cite{ivanov2015}: 

\begin{equation} \label{38}
\mathbf{H}_k =
\begin{bmatrix}
    2W_{\hat{\boldsymbol{b}}^{b}_{\text{model}}} & \mathbf{0}_{3\times3} \\
    2W_{\boldsymbol{r}^{b}_{\text{model}}} & \mathbf{0}_{3\times3}
\end{bmatrix}
\end{equation}

Here, $W_{\hat{\boldsymbol{b}}^{b}_{\text{model}}}$ and $W_{\boldsymbol{r}^{b}_{\text{model}}}$ denote the skew-symmetric matrix for $\hat{\boldsymbol{b}}^{b}_{\text{model}}$ and $\boldsymbol{r}^{b}_{\text{model}}$, respectively.

$\mathbf{R}$ constitutes the standard deviation for both the sensors' noise, given as:

\begin{equation} \label{39}
    {\mathbf{R} = \begin{bmatrix}
        \mathrm{\sigma_{mtm}^{2}}\mathbf{I_{3\times3}} & \mathbf{0_{3\times3}} 
        \\   \mathbf{0_{3\times3}} & \mathrm{\sigma_{ss}^{2}}\mathbf{I_{3\times3}}
    \end{bmatrix}}
\end{equation}

$\sigma_{mtm}$ is the standard deviation for the magnetometer noise and $\sigma_{ss}$ is the standard deviation for the sun sensor noise.

\section{Simulation Results}
\label{Sec3}
The analysis and comparison of the geomagnetic field models on satellite ADCS is done upon the specifications of Skoltech-F CubeSat, as mentioned in \cite{bano2024development}.

The satellite inertia tensor is given as:
\begin{equation} \label{40}
\mathbf{J} = 
\begin{bmatrix}
0.05466 & -0.00004 & -0.00006 \\
-0.00004 & 0.05531 & 0.00029\\
-0.00006 & 0.00029 & 0.01201
\end{bmatrix} kg.m^2
\end{equation}

The satellite orbit is taken be sun-synchronous, which an inclination of 97 degrees. The environmental and control parameters for the simulation setting are given in Tables \ref{table1} and \ref{table2} as:

\begin{table}[h!]
\center
\caption{\label{table1} Environmental Parameters}
\begin{tabular}{cc}
\toprule
Disturbance Torque Deviation (Nm) & $3 \times 10^{-7}$ 
\\
\midrule
Magnetic Field Deviation (T)  &  $2 \times 10^{-6}$ \\
\bottomrule
\end{tabular}
\end{table}

\begin{table}[h!]
\center
\caption{\label{table2} Control Parameters}
\begin{tabular}{cc}
\toprule
Required Quaternion & $[1, 0, 0, 0]^{T}$ \\
\midrule Proportional Gain & 0.115 \\
\midrule
Derivative Gain & 0.245 \\
\bottomrule
\end{tabular}
\end{table}

\subsection{Magnetometer Modeling}
The standard deviation $\sigma_{mtm}$ for the magnetometer is taken to be $1 \times 10^{-7}$ T. 
\newpage
\begin{figure}[hbt!]
\centering
\centerline{\includegraphics[width=19pc]{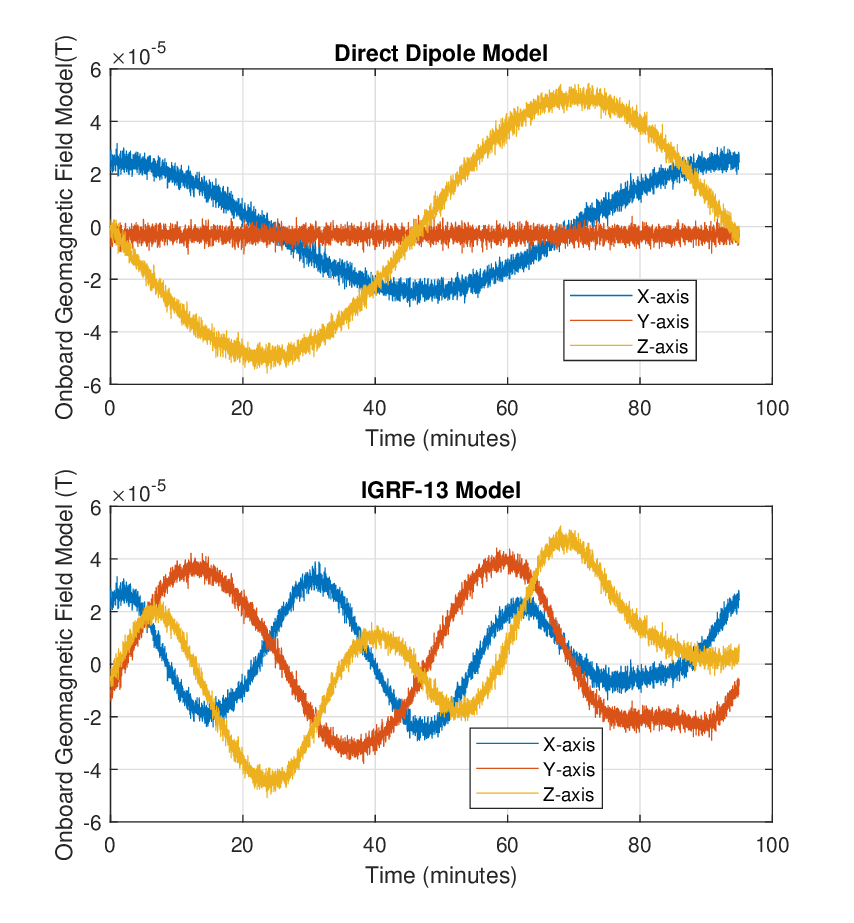}}
\caption{Geomagnetic Field Modeling}
\label{Fig4}
\end{figure}

Figure \ref{Fig4} shows the onboard magnetic field of the Earth with environmental noise deviation, using Direct Dipole model and IGRF-13 model. It is evident that the Direct Dipole model is quite simple and treats the Earth’s magnetic field as a dipole while, the IGRF model accurately models all the sources of the geomagnetic. Moreover, the IGRF model is specific to changes in the magnetic field with time, while the Direct Dipole models the magnetic field as simple trigonometric functions, independent of time.

\subsection{Orientation Stability}
The specifications for the sun sensors and reaction wheels used in this research are similar to the ones used in Skoltech-F CubeSat, given in \cite{bano2024development}. For the EKF, the process and measurement noise covariance matrices are formed based on the disturbance and measurement deviations mentioned above. The initial state estimate $\boldsymbol{x_o^+}$ for the quaternions is determined by the TRIAD model \cite{gokcay2022triad} and for the angular rates is taken to be 0. The initial state error covariance matrix $\mathbf{P_{o} = P_o^+}$ is given as:
\begin{equation} 
\mathbf{
    {P_{o} }= \begin{bmatrix}
        \sigma^2_{q_{o}}\mathbf{I_{3\times3}}  & \mathbf{0_{3\times3}}
        \\   \mathbf{0_{3\times3}} & \sigma^2_{\omega_{o}}\mathbf{I_{3\times3}}
    \end{bmatrix}}
    = 
    \begin{bmatrix}
       1^{2}\mathbf{I_{3\times3}}  & \mathbf{0_{3\times3}}
        \\   \mathbf{0_{3\times3}} & 0.1^{2}\mathbf{I_{3\times3}}
    \end{bmatrix}
\end{equation}

\subsubsection{Implementation of Direct Dipole Model}
In this subsection, the magnetometer measurements for attitude determination are generated using the Direct Dipole model for the geomagnetic field model. 

\begin{figure}[hbt!]
   \centerline{\includegraphics[width=22pc]{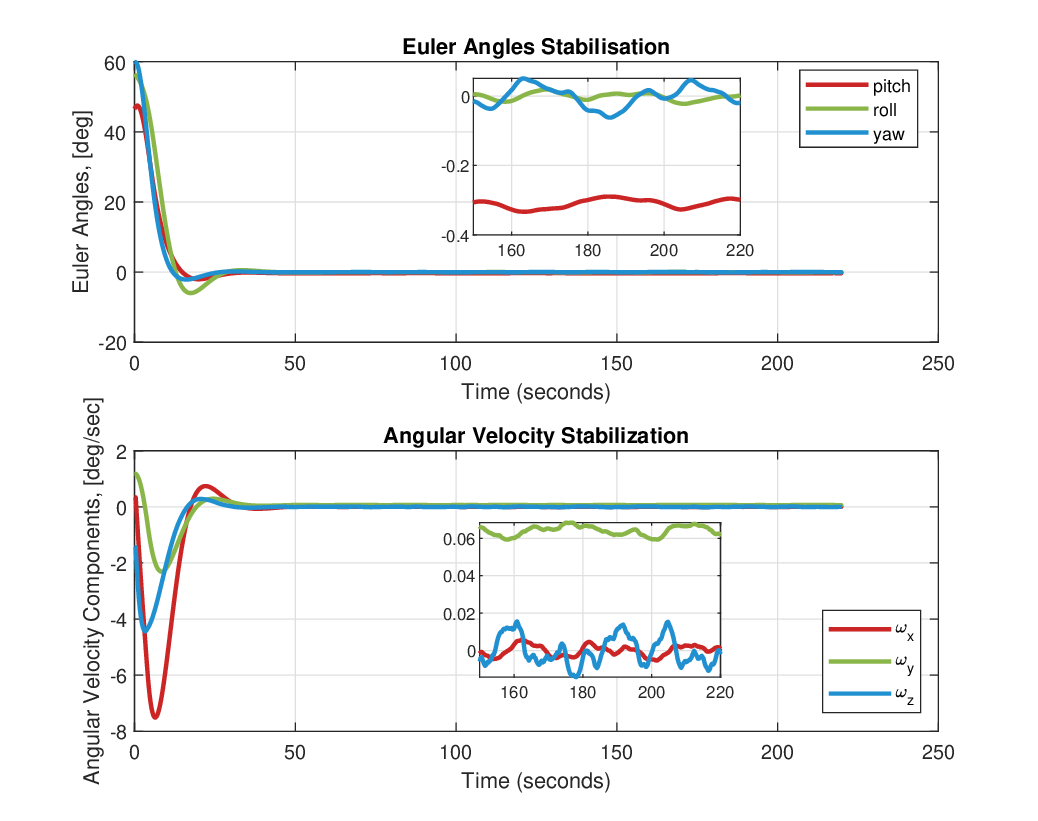}}
    \caption{Attitude Stabilization}
    \label{Fig5}
\end{figure}

Figure \ref{Fig5} above shows satellite states stabilization within acceptable limits achieved in less than a minute. For angles, the error is less than 0.4 degrees, and for rates, the error is less than $1.4\times10^{-4}$ deg./sec. 

\begin{table}[h!]
\caption{\label{table3} RMSE (Root Mean Square Errors)}
\center
\begin{tabular}{ccc}
\toprule
Parameter & RMSE \\
\midrule
Pitch (degrees) &  0.308 \\

Roll (degrees) &  0.01 \\

Yaw (degrees)&  0.03 \\
\midrule
{$\omega_{x}$ (deg/sec)} &  {$4.5\times10^{-5}$} \\

{$\omega_{y}$ (deg/sec)} &  {$4.2\times10^{-5}$} \\

{$\omega_{z}$ (deg/sec)} &  {$1.3\times10^{-4}$} \\
\bottomrule
\end{tabular}
\end{table}

Table \ref{table3} shows the RMSE of the Euler angles and angular rates. 

\subsubsection{Implementation of IGRF-13 Model}
In this subsection, the magnetometer measurements for attitude determination are generated using the IGRF-13 model for the geomagnetic field. 

\begin{figure}[hbt!]
\centerline{\includegraphics[width=22pc]{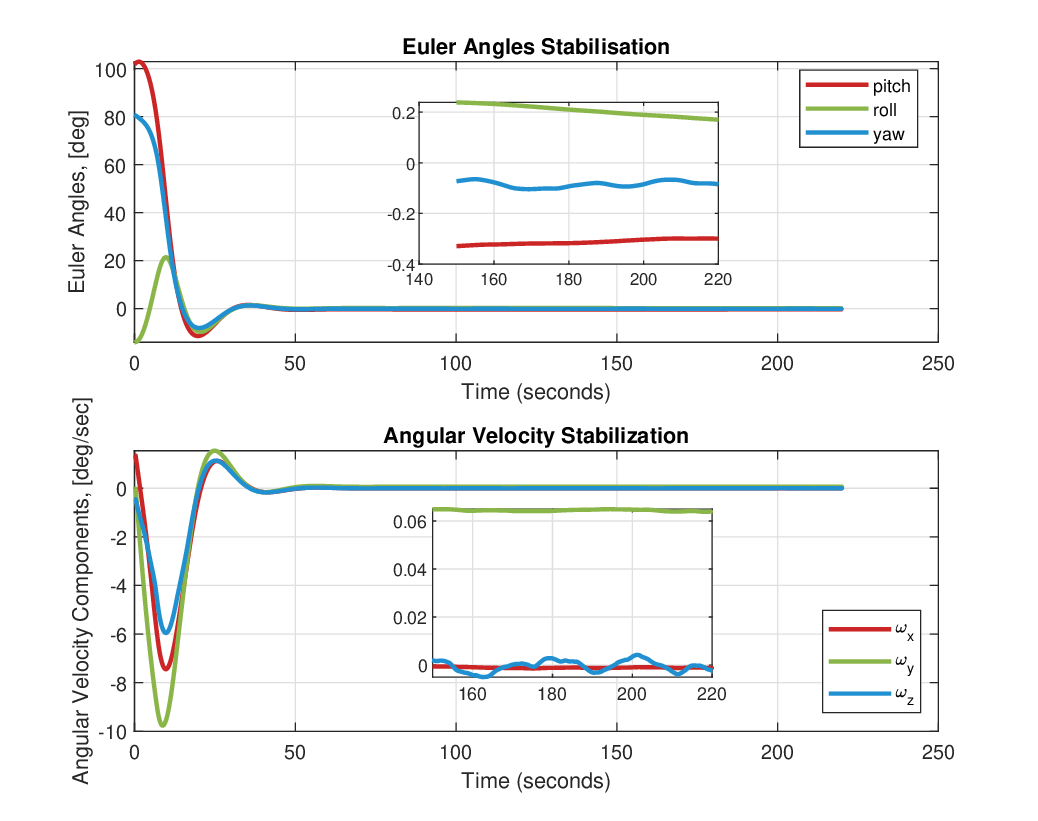}}
\caption{Attitude Stabilization}
\label{Fig6}
\end{figure}
\newpage
Figure \ref{Fig6} above shows satellite states stabilization within acceptable limits achieved in less than a minute. For angles, the error is less than 0.4 degrees, and for rates, the error is less than $4\times10^{-5}$ deg./sec. Moreover, it can be seen from Figures \ref{Fig5} and \ref{Fig6} that the angular rate about the y-axis stabilizes around 0.06 deg/sec, which is the mean motion of the satellite in the orbit.

\begin{table}[h!]
\caption{\label{table4} RMSE (Root Mean Square Errors)}
\center
\begin{tabular}{cc}
\toprule
Parameter & RMSE \\
\toprule
Pitch (degrees) &  0.31 \\

Roll (degrees) &  0.2 \\

Yaw (degrees)&  0.09 \\
\midrule
{$\omega_{x}$ (deg/sec)} &  {$2\times10^{-5}$} \\

{$\omega_{y}$ (deg/sec)} &  {$3.1\times10^{-5}$} \\

{$\omega_{z}$ (deg/sec)} &  {$3.5\times10^{-5}$} \\
\bottomrule
\end{tabular}
\end{table}

Table \ref{table4} shows the RMSE of the Euler angles and angular rates. 
\subsubsection{Discussion}
The results above demonstrate the effectiveness of utilizing both the Direct Dipole and IGRF-13 Model for magnetometer measurements in stabilizing the satellite's orientation. While the RMSE for Euler angles slightly favors the Direct Dipole model, the IGRF model yields slightly lower RMSE for angular rates. However, both models perform well within the acceptable range, showcasing no significant deviation. Thus, considering computational efficiency, the Direct Dipole model emerges as the preferred choice. This is because it meets the angular accuracy requirement of the Skoltech-F mission to align the orbital and body frame within 1 degree.

\subsection{Unloading of Reaction Wheels}
Unloading of the reaction wheels is done using an array of three magnetorquers in XYZ configuration. The specifications of the magnetorquer is given as:

\begin{table}[h!]
\caption{\label{table5} Magnetorquer Specifications}
\center
\begin{tabular}{cc}
\toprule
Area of the coil  $\mathbf({m^2})$ & 0.0025
\\ \midrule
Number of turns & 200 
\\ \midrule 
Wire Resistance $(\Omega)$  & 25  \\ \hline
Voltage from EPS (V) & 5
\\
\bottomrule
\end{tabular}
\end{table}

\subsubsection{Implementation of Direct Dipole Model}
In this subsection, the geomagnetic field modeling for the B-dot controller is done through Direct Dipole model.

\begin{figure}[hbt!]
\centerline{\includegraphics[width=21pc]{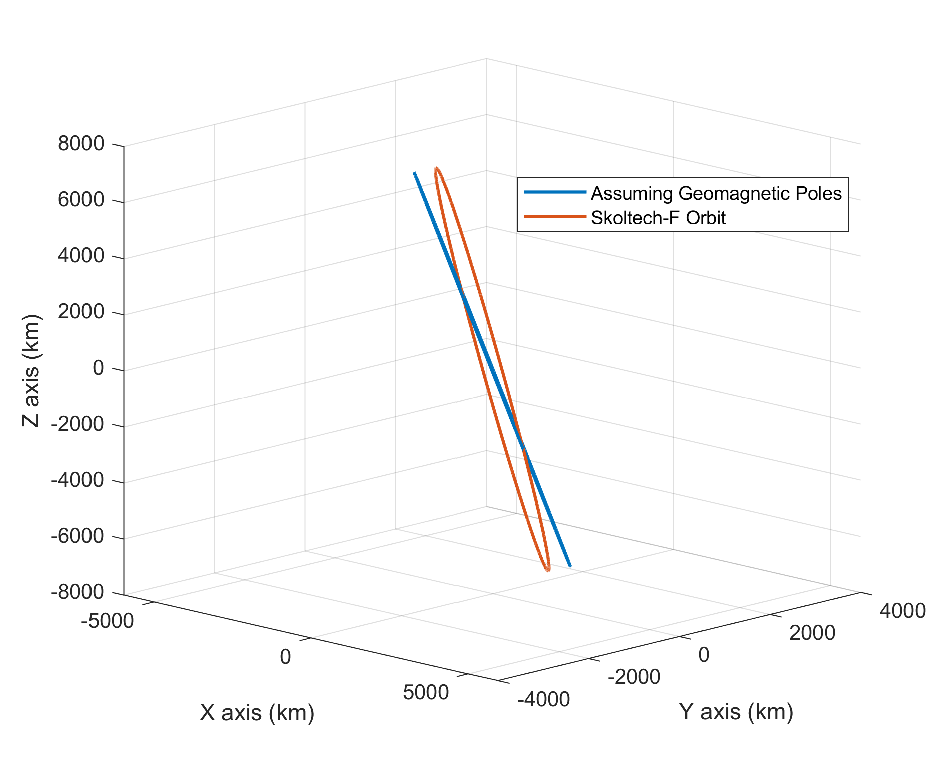}}
\caption{Interaction between Satellite Orbit and Magnetic Field Poles in Direct Dipole Model}
\label{Fig7}
\end{figure}

The Direct Dipole model does not account for the slight rotation of the magnetic field axis. Figure \ref{Fig7} analyzes the interaction points between the fixed magnetic axis and the satellite orbit, crucial since magnetic control is ineffective near the geomagnetic poles and equator \cite{vallado2013fundamentals}. The portion of the orbit farthest from the geomagnetic axis is identified to optimize the B-dot controller for reaction wheel unloading.

\begin{figure}[hbt!]
\centerline{\includegraphics[width=22pc]{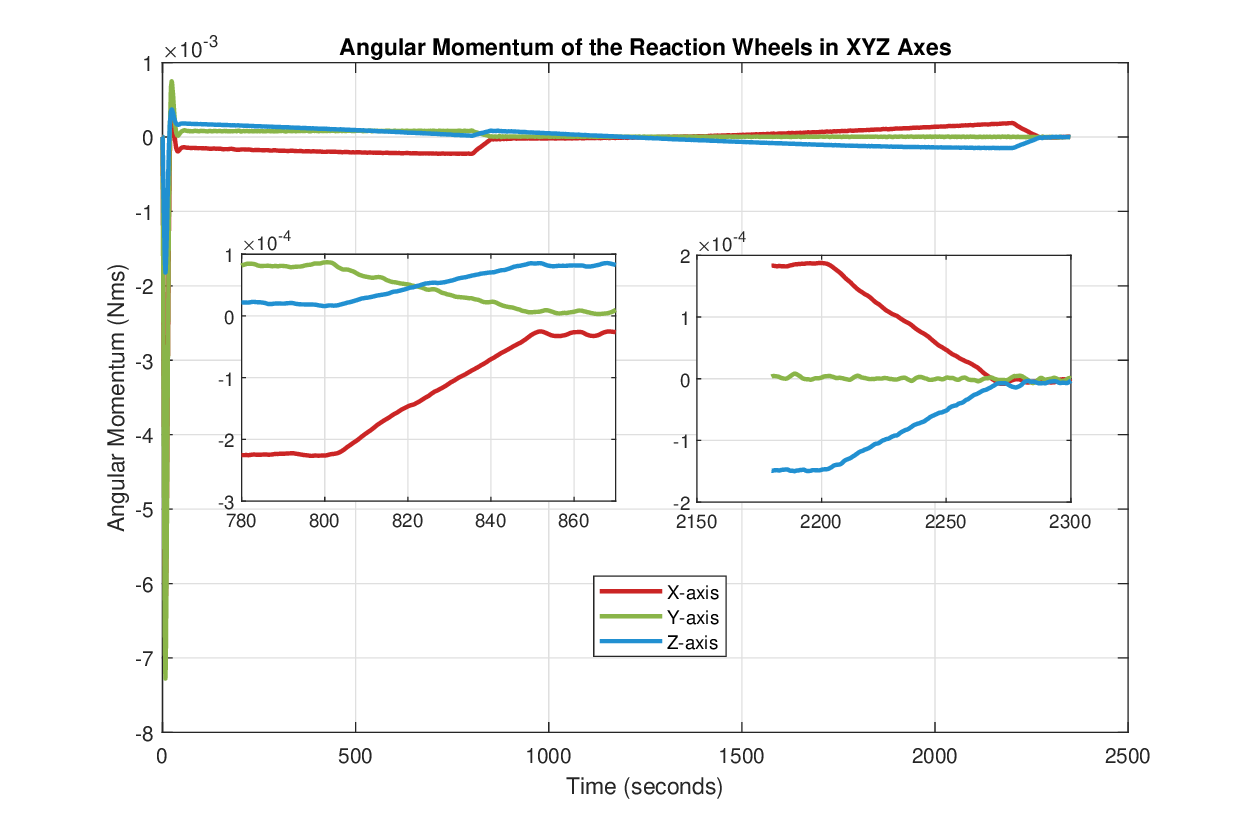}}
\caption{Unloading of Reaction Wheels Angular Momentum}
\label{Fig8}
\end{figure}

Figure \ref{Fig8} above shows unloading of the reaction wheels angular momentum done with the help of magnetorquers. The unloading is done in two intervals with each interval unloading two axes at a time. 

\begin{table}[h!]
\caption{\label{table6}Angular Momentum of Reaction Wheels}
\center
\begin{tabular}{ccc}
\toprule
Time (sec)  & Axis  & Magnitude (Nms) 
\\
\midrule
\multirow{3}{*}{800} &  X     &   $-2.26 \times 10^{-4}$        \\  
                    & Y &    $0.87 \times 10^{-4}$       \\  
                    &  Z     &  $0.16 \times 10^{-4}$         \\ \hline
\multirow{3}{*}{850} &  X     &   $-0.3 \times 10^{-4}$        \\  
                    & Y &    $0.06 \times 10^{-4}$         \\  
                    &  Z     &  $0.9\times10^{-4}$   \\ \hline
\multirow{3}{*}{2200} &  X     &  $1.9\times 10^{-4}$              \\  
                    & Y &    $0.2\times 10^{-4}$         \\  
                    &  Z     &  $-1.5\times 10^{-4}$       \\ \hline       
\multirow{3}{*}{2270} &  X     & $ -0.05\times 10^{-4}$              \\  
                    & Y &    $-0.007\times 10^{-4}$         \\  
                    &  Z     &  $-0.09\times 10^{-4}$       \\ \hline                    
\end{tabular}
\end{table}

Table \ref{table6} shows significant decrease in angular momentum of the reaction wheels
after both the unloading intervals. It can be observed that the angular momentum of the wheels decrease by about 97\% about the X-axis, about 98\% about the Y-axis and about 44\% about the Z-axis. Hence, the unloading algorithm successfully dissipates the angular momentum effectively.

\subsubsection{Implementation of IGRF Model}
In this subsection, the geomagnetic field modelling for the B-dot controller is done through IGRF model. 

\begin{figure}[hbt!]
\centerline{\includegraphics[width=22pc]{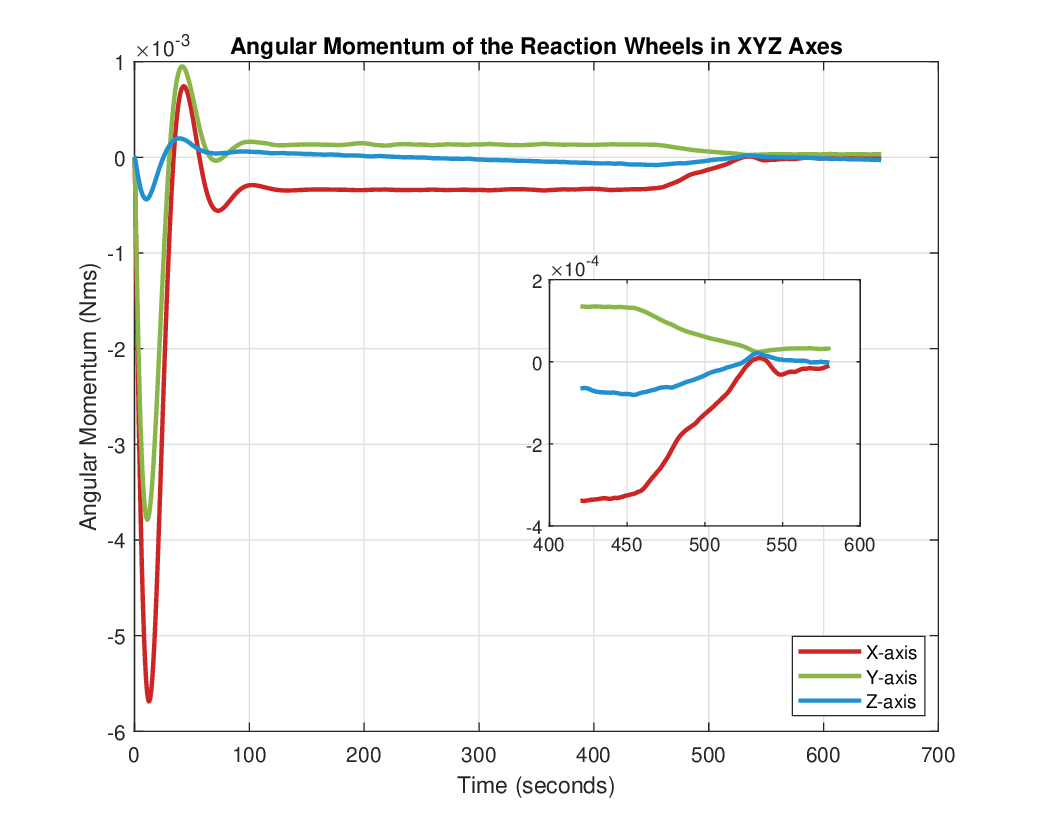}}
\caption{Unloading of Reaction Wheels Angular Momentum}
\label{Fig9}
\end{figure}
As evidenced from Figure \ref{Fig9}, IGRF-13 model is quite accurate in modeling geomagnetic field because it compensates for all the geomagnetic sources as well as accounts for the Earth's rotation. Hence, it is possible to perform unloading of the angular momentum about all the three axes of the reaction wheels with this model.

\begin{table}[h!]
\caption{\label{table7}Angular Momentum of Reaction Wheels}
\center
\begin{tabular}{ccc}
\hline
Time (sec)  & Axis  & Magnitude (Nms) 
\\
\hline
\multirow{3}{*}{450} &  X     &  $-3.2 \times 10^{-4}$         \\  
                    & Y &    $1.3 \times 10^{-4}$       \\  
                    &  Z     &   $-0.8 \times 10^{-4}$      \\ \hline
\multirow{3}{*}{560} &  X     &   $-0.2\times 10^{-4} $      \\  
                    & Y &   $0.3\times 10^{-4}$           \\  
                    &  Z     &  $ -0.01\times 10^{-4} $              \\ \hline               
\end{tabular}
\end{table}

Table \ref{table7} above shows significant decrease in angular momentum of the reaction wheels after a single unloading interval. It can be observed that the angular momentum of the wheels decrease by about 93\% about the X-axis, about 77\% about the Y-axis and about 98\% about the Z-axis. Hence, the unloading algorithm successfully dissipates the angular momentum effectively. 

\subsubsection{Discussion}
The geomagnetic field generation for the unloading algorithm is implemented by both Direct Dipole model and IGRF-13 model. It can be seen that percentage decrease in built-up angular momentum of the reaction wheels by Direct Dipole model is slightly more than using the IGRF model. However, this is because the Direct Dipole model requires two separate intervals for implementing the unloading algorithm, while the IGRF model performs it in within a single interval. 

Hence, in this analysis the IGRF-13 model is preferred as it accurately models the magnetic field, and adequately reduces the angular momentum of the reaction wheels within the required range. Another reason for choosing this model is that performing unloading within a single interval reduces the chance of excessive angular momentum zero crossing, which might affect the performance of the reaction wheels in the long run. 

\section{Conclusion}
\label{Sec4}
This research focuses on the comparison of two geomagnetic field models (Direct Dipole Model and IGRF-13 model) and the analysis on their behaviors in different ADCS modes. The Direct Dipole Model proved to be preferable for sensor modeling of the magnetometer due to its computational efficiency and comparable orientation accuracy. The angular RMSE resulted in this case was less than a degree. These errors can further be decreased by implementing advanced control techniques such as Model Predictive Control (MPC), Sliding Mode Control (SMC), etc. and by employing a more efficient torque decomposition algorithm for the reaction wheels.

For the momentum dumping of reaction wheels, unloading algorithm was implemented using both these geomagnetic field models. The unloading interval (significant aspect of an unloading algorithm) was derived for the Direct Dipole Model analysis since it assumes a fixed tilt of the magnetic field axis (i.e., it does not rotate with the Earth). The IGRF-13 model was found to be more suitable for modeling the magnetic field for magnetorquers in B-dot control, optimizing the unloading of reaction wheels and avoiding zero crossing. 

This study, implemented on the Skoltech-F CubeSat mission, demonstrates the practical application of selected geomagnetic field models for various ADCS modes by integrating sensors, actuators and estimation algorithms. The findings provide valuable guidance for choosing appropriate models based on mission-specific requirements and computational constraints. Future studies could extend this comparison to different satellite classes, such as FemtoSats, where computational resources are highly limited, or larger satellites that demand higher accuracy. Additionally, incorporating space weather effects into geomagnetic field modeling is recommended, as they can significantly influence magnetometer readings and eventually impact attitude determination accuracy.

\section*{References}

\end{document}